\documentclass[conference]{IEEEtran}
\IEEEoverridecommandlockouts
\usepackage{cite}
\usepackage{amsmath,amssymb,amsfonts}
\usepackage{algorithm}
\usepackage{algpseudocode}
\usepackage{graphicx}
\usepackage{hyperref}
\usepackage{textcomp}
\usepackage{xcolor}
\usepackage{subcaption}
\usepackage{booktabs}
\usepackage{pifont}
\usepackage{comment}
\newcommand{\cmark}{\ding{51}}%
\newcommand{\xmark}{\ding{55}}%
\def\BibTeX{{\rm B\kern-.05em{\sc i\kern-.025em b}\kern-.08em
    T\kern-.1667em\lower.7ex\hbox{E}\kern-.125emX}}

\begin{document}

\title{A Tip for IOTA Privacy: IOTA Light Node Deanonymization via Tip Selection}

\author{\IEEEauthorblockN{Hojung Yang}
\IEEEauthorblockA{\textit{School of Cybersecurity} \\
\textit{Korea University}\\
Seoul, Korea \\
ghwjd0816@korea.ac.kr}
\and
\IEEEauthorblockN{Suhyeon Lee}
\IEEEauthorblockA{\textit{Tokamak Network} \\
suhyeon@tokamak.network}
\and
\IEEEauthorblockN{Seungjoo Kim}
\IEEEauthorblockA{\textit{School of Cybersecurity} \\
\textit{Korea University}\\
Seoul, Korea \\
skim71@korea.ac.kr}
}

\maketitle

\begin{abstract}
IOTA is a distributed ledger technology that uses a Directed Acyclic Graph (DAG) structure called the Tangle.
It is known for its efficiency and is widely used in the Internet of Things (IoT) environment.
Tangle can be configured by utilizing the tip selection process.
Due to performance issues with light nodes, full nodes are being asked to perform the tip selections of light nodes.
However, in this paper, we demonstrate that tip selection can be exploited to compromise users' privacy.
An adversary full node can associate a transaction with the identity of a light node by comparing the light node's request with its ledger.
We show that these types of attacks are not only viable in the current IOTA environment but also in IOTA 2.0 and the privacy improvement being studied.
We also provide solutions to mitigate these attacks and propose ways to enhance anonymity in the IOTA network while maintaining efficiency and scalability.
\end{abstract}

\begin{IEEEkeywords}
Cryptocurrency, IOTA, Tangle, DAG, Privacy, Anonymity, Blockchain
\end{IEEEkeywords}

\section{Introduction}
IOTA uses a directed acyclic graph (DAG) structure called the Tangle, instead of the Nakamoto blockchain's linked list structure\cite{ref1}.
Due to this unique structure, IOTA offers advantages in scaling and transaction speed, making it ideal for large Internet of Things (IoT) environments\cite{ref8,ref9}.
However, the special structures pose challenges related to security and privacy\cite{ref5}.

IOTA introduced a centralized component called the \textit{coordinator} to maintain stability and security.
The coordinator technology ensures network stability by validating and approving transactions.
However, the existence of such coordinators goes against decentralization, one of the core values of distributed ledger technology\cite{ref3}.

One of the core technologies of IOTA is the tip selection algorithm, which plays a crucial role in selecting and validating transactions on the IOTA network.
This algorithm can cause performance issues on resource-constrained devices that act as light nodes\cite{ref10}.
As a result, light nodes rely on full nodes to perform the tip selection process.

\begin{figure}[bt]
    \centering
    \includegraphics[width=\linewidth]{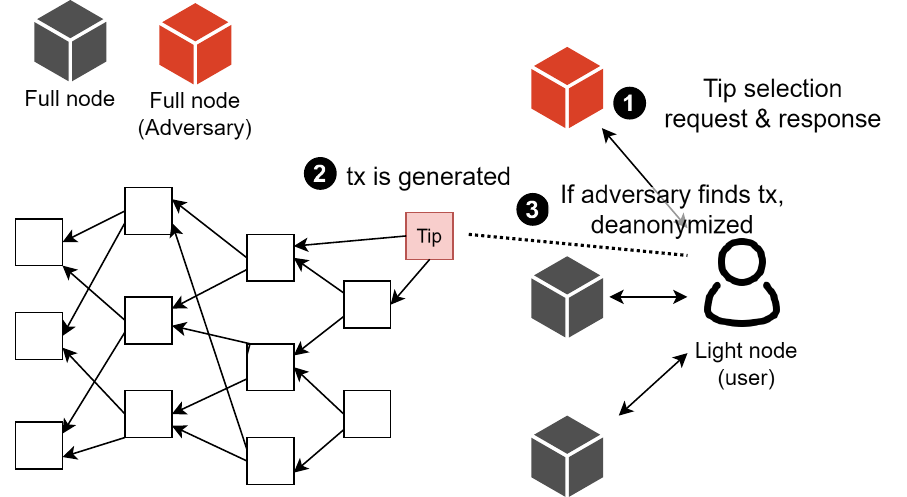}
    \caption{If a light node requests tip selection to an adversarial node, the attacker can determine the identity probabilistically after the transaction is generated.}
    \label{fig: mainmodel}
    \vspace{-0.5cm}
\end{figure}

Our paper introduces a new privacy deanonymization attack that targets the IOTA tip selection algorithm. Figure \ref{fig: mainmodel} illustrates the primary concept of the attack.
The attack is carried out by comparing the ledger with the tip selection results provided by the adversarial full node to the light node, thus creating a link between the two.
What sets this attack apart from known cryptocurrency deanonymization attacks is that it can be applied to each transaction to cause significant damage with fewer resources.
We assessed the effectiveness and triumph of the attack by measuring its level of anonymity in various settings.
Furthermore, we introduce a mitigation to address this attack.
The mitigation we provide focuses on the tip selection itself and suggests a new direction for improving the anonymity of the IOTA network.

The contributions of this paper are as follows:

\begin{itemize}
    \item We proposed a novel deanonymization technique in IOTA using the light node's tip selection mechanism.
    \item We performed a comprehensive analysis of our technique on various IOTA environments including the real world node distribution, decentralized node distribution, privacy-enhanced environment, and IOTA 2.0.
    \item We investigated possible mitigations to improve privacy for IOTA light nodes against our deanonymization.
\end{itemize}

The rest of the paper is organized as follows:
Section \ref{section: related works} introduces related studies about IOTA focused on tip selection and privacy.
Section \ref{section: system model} introduces the system model we set and our approach to evaluating tip selection anonymity, followed by a detailed analysis in Section \ref{section: analysis}.
We present possible mitigation for our attack in Section \ref{section: mitigation}.
Section \ref{section: discussion} outlines the significance and limitations of our approach and suggests potential future work.
Finally, Section \ref{section: conclusion} concludes the paper.

\section{Related Works}
\label{section: related works}
This section reviews the overall structure, tip selection algorithm, and privacy aspects of the IOTA network.

\subsection{IOTA network}

In contrast to the blockchain's successive block structures, the IOTA Tangle stores data in each node referencing previous transactions on a graph.
Therefore unlike the blockchain, mining operations or miners are not required.
The key features of the Tangle structure include\cite{ref11}:
\begin{itemize}
    \item Transaction speed: The DAG structure speeds up transaction processing. This is especially effective when there are more participants in the network.
    \item Scalability: IOTA's DAG is scalable in large environments such as IoT and can work efficiently in resource-constrained environments such as wireless sensor networks.
\end{itemize}

The concept of coordinators was introduced to ensure the initial stability and security of the IOTA network\cite{ref12}.
Coordinators are responsible for validating and approving transactions centrally.
This approach is used to improve initial network performance and enhance security and aims to operate safely in distributed environments in the future.
However, the existence of a coordinator shows that IOTA is not fully distributed, resulting in some centralization\cite{ref3}.
IOTA is making efforts such as developing a \textit{coordicide}, replacing a coordinator in IOTA 2.0, to solve this problem\cite{url1}.

There are some main areas of research on IOTA.
There is ongoing research on tip selection to improve the performance and stability of IOTA.
There are also many studies focusing on the anonymity aspect of IOTA.

\subsection{Tip selection}
\label{subsection: tip selection}
IOTA's tip selection addresses key questions about selecting transactions from the distributed ledger and adding new transactions appropriately\cite{ref1,ref13}.
The algorithm plays a key role in IOTA's Tangle structure.
It determines which transactions should be selected, validated, and added from those not yet in the distributed ledger.
This process enables fast transaction processing and secure distributed ledger operations and plays an important role in large-scale transaction environments such as IoT-based applications.

When it comes to performing tip selection, light nodes may face difficulties due to computational issues\cite{ref10}.
In such cases, these nodes can ask for help from the surrounding full nodes.
It is recommended to request tip selection from multiple full nodes to enhance privacy.

There has been research into faster, lighter, and more efficient tip selection.
Wang et al.\cite{ref19} introduced dynamic tip selection, which improves the anti-splitting ability of IOTA by selecting according to the situation of the distributed ledger instead of using a single selection algorithm.
Xiao et al.\cite{ref20} introduced a tip selection probability distribution precalculation method so that the tip selection phase can proceed faster. 
Kusmierz et al.\cite{ref21} conducted a comparative analysis of various tip selection algorithms in terms of stability, overhead, and performance.

\subsection{Privacy}
\label{subsection: privacy}
IOTA's privacy issues are active research areas\cite{ref17,ref5}.
For privacy issues, researchers have focused on distributed ledger analysis attacks and defenses.

Tennant\cite{ref5} theoretically analyzed privacy issues in IOTA.
The author introduced that there are conceptually four privacy issues in IOTA: address reuse, taint analysis, transaction analysis, and metadata analysis.
In the case of address reuse, unlike account-based cryptocurrencies such as Ethereum, IOTA can generate any number of new addresses, so it is resistant if the wallet supports the automatic generation of new addresses.
On the other hand, taint analysis, transaction analysis, and metadata analysis are attacks that can be performed on IOTA.
Especially for metadata analysis, peer discovery on IOTA is manual and requires a static IP, making it difficult to route through anonymity techniques.

The author also proposed methodologies to enhance privacy such as a centralized mixer and ring signature.
The proposed methodologies primarily focus on aspects related to taint analysis and the topology of full or light nodes.
Some of these methodologies have been thoroughly researched to tailor them to the IOTA\cite{ref17,url4}.
To counter taint analysis, centralized mixers\cite{url4,url5} and coin swap\cite{ref18} can be used to break the association between pairs of addresses on the IOTA network.

Sarfraz et al.\cite{ref17} analyzed the applicability of the previously proposed methodology to IOTA.
They found that both Centralized Mixers\cite{url4,url5} and Coin Swap\cite{ref18} can be adapted to the IOTA network.
In addition, they presented a fully decentralized solution named \textit{decentralized mixing}, which uses a digital multi-signature scheme and decryption mix-nets.
However, their solution has the disadvantage of increasing computational overhead as the number of participants increases.
Nonetheless, they demonstrated that an adversary cannot forge signatures by stealing portions of a participant's private key.

One approach involves utilizing cross-chain bridges to transfer assets to different cryptocurrencies, followed by leveraging mixers to enhance privacy.
This method draws inspiration from similar techniques used in other blockchain networks like Ethereum and Bitcoin.
For example, Ethereum's Tornado Cash\cite{tornado} is known for obfuscating transaction history through mixer contracts.
Similarly, CoinJoin\cite{coinjoin} protocols are used in Bitcoin to achieve the same goal. 
By adopting and adapting these techniques to the IOTA ecosystem, it becomes possible to improve privacy.

Our research takes a unique approach to anonymity in IOTA, as compared to previous studies.
Instead of analyzing the ledger to uncover associations between addresses, we establish a direct link between users and addresses.
Moreover, our research emphasizes enhancing privacy rather than the efficiency of tip selection.

\section{System Model and Approach}
\label{section: system model}
This section introduces our system model and approach for evaluating anonymity.

\subsection{System model}
In this paper, we assume that a light node does not perform a tip selection directly, but rather requests a tip selection from multiple full nodes.
As mentioned in Section \ref{subsection: tip selection}, light nodes are not directly connected to the Tangle but are connected to full nodes and utilize their resources to attach transactions to the Tangle.
We also assume that each full node will send a distinguishable tip selection response.

We follow the steps described in Algorithm \ref{alg} in our attack.
Firstly, the light node $L$ requests the full node to process its tip selection.
Once $L$ receives the result, it registers a transaction in the ledger based on it. 
At this point, the adversary node checks the ledger.
If the location of the registered transaction matches the tip selection result that it provided, it confirms that the transaction was made by $L$.

\begin{algorithm}[b]
\caption{Deanonymization Attack}
\label{alg}
\renewcommand{\algorithmicrequire}{\textbf{Input:}}
\renewcommand{\algorithmicensure}{\textbf{Output:}}
\begin{algorithmic}[1]
    \Require Light node $L$, full nodes set $F$, adversary nodes set\par \hskip-0.5em $C \subset F$
    \Ensure Pair of L's address and network configuration\par\hskip0.3em $\langle A_L, N_L\rangle$
    \State $L$ requests tip selection from $F' \subset F$
    \State $F'$ responds, $F' \wedge C$ sends unique responses
    \State $L$ sends a transaction based on the response
    \State Analyze ledger to update $\langle A_L, N_L\rangle$ if match found with $F' \wedge C$'s response
    \State \Return $\langle A_L, N_L\rangle$

\end{algorithmic}
\end{algorithm}

Our analysis begins by examining the decentralized environment, which serves as the basis for all our subsequent analyses.
In the decentralized environment, full nodes are evenly distributed, and light nodes can request tip selection from any of the full nodes.
Once we have analyzed the decentralized environment, we move on to investigate three distinct situations.
The first one is the real world, where we explore the issue of requesting tip selection from a neighboring full node based on the actual distribution of full nodes.
Secondly, we examine the privacy-enhanced environment, where we employ the methodologies elaborated in Section \ref{subsection: privacy}.
Finally, we analyze the situation after the implementation of IOTA 2.0.

\subsection{Approach}
Díaz et al. \cite{ref7} provide a formal definition for the degree of anonymity as the following.
Firstly, let $H(X)$ be the Shannon entropy of the system after an attack has taken place with the number of possible senders, $N$, and the probability, $p_i$,
we can calculate $H(X)$ as
\begin{equation}
    H(X) = -\sum_{i=1}^{N}p_i log_2(p_i) .
\end{equation}
Let $H_m$ be the maximum state of entropy in a system, wherein every entity is equally likely to be the sender.
\begin{equation}
    H_m = log_2(N)
\end{equation}
The information the attacker has learned with the attack can be expressed as $H_{m}$$-$$H(X)$.
After we divide it by $H_m$ for normalization, the degree of anonymity is defined as:
\begin{equation}
    d = \frac{H(X)}{H_m}
\end{equation}

The anonymity of cryptocurrencies has been analyzed using Shannon entropy.
It is a method that helps to determine the success rate and intensity of attacks on cryptocurrencies and to evaluate the applied methodology's effectiveness.
To calculate the anonymity of the P2P network scheme and Ethereum account linkage, Shannon entropy was used by Sharma et al. \cite{ref2} and Béres et al. \cite{ref16}, respectively.
Biryukov et al.\cite{ref15} determined the attack success rate by computing the anonymity degree using Shannon entropy.
Sarfraz et al.\cite{ref17} used the same approach to calculate the anonymity of the IOTA ledger with decentralized mixing.

Our approach involves identifying the corresponding transaction originator by comparing the request and response of tip selection with the message created in the ledger.
This information content is then used to quantify the probability of a particular event.
Table~\ref{table_not} shows the basic notation and definition of variables used in anonymity analysis.

\begin{table}[tb]
\caption{Basic notation and definition of variables}
\label{table_not}
\centering
\setlength{\tabcolsep}{5pt}
\renewcommand{\arraystretch}{1.5}
\begin{tabular}{c p{7cm}}
\textbf{Symbol} & \multicolumn{1}{c}{\textbf{Description}} \\ \hline
\multicolumn{1}{c}{$N$} &  The total number of full nodes in the network. \\
\multicolumn{1}{c}{$C$} &  The total number of adversary nodes in the network. \\
\multicolumn{1}{c}{$B_i$} &  An event where light node $i$ requests tip selection. \\
\multicolumn{1}{c}{$A$} &  An event where the adversary node receives tip selection. \\
\multicolumn{1}{c}{$M$} &  The number of full nodes the light node requests for tip selection.  \\
\multicolumn{1}{c}{$N_s$} &  The random variable representing the number of adversary nodes requested by a light node for tip selection.\\
\multicolumn{1}{c}{$P(A|B_i)$} &  The conditional probability of an adversary node receiving a tip selection query from light node $i$. \\
\multicolumn{1}{c}{$P(N_s)$} &  The probability that $N_s$ adversary node will be requested tip selection.
\end{tabular}
\end{table}

If event $A$ occurs, given that $B_i$ has already occurred, an adversary can distinguish which light node created the message. To calculate the probability of $A$ under the condition $B_i$, we first calculate the probability of random variable $N_s$. When $N_s$ adversary nodes are selected in $B_i$, $N_s/M$ is the probability that light node $i$ follows adversaries' tip selection.\\
To calculate the probability of a light node being deanonymized, denoted by $P(A|B_i)$, the following formula is used:
\begin{equation} \label{eq. 4}
    P(A|B_i)=\sum_{k=0}^{min(M,C)}\frac{k\cdot P(N_s)}{M}
\end{equation}

Since $P(N_s)$ is the probability of $N_s$ successes in $M$ draws
without replacement, from a full node of size $N$ that contains exactly $C$ objects,
$N_s$ follows hyper-geometric distribution.
\begin{equation} \label{eq. 5}
    P(N_s)=\frac{\binom{C}{N_s} \cdot \binom{N-C}{M-N_s}} {\binom{N}{M}}
\end{equation}

From Eq. \ref{eq. 4} and Eq. \ref{eq. 5}, $P(A|B_i)$ is reduced to 
\begin{equation}
    P(A|B_i)=\sum_{k=0}^{min(M,C)}\frac{k}{M}\cdot \frac{\binom{C}{N_s} \cdot \binom{N-C}{M-N_s}} {\binom{N}{M}} .
\end{equation}

\section{Analysis on Anonymity in IOTA Tip Selection}
\label{section: analysis}
This section focuses on the analysis of anonymity in tip selection on the IOTA network.
We verify the validity of our attack in three different environments: The decentralized environment, the real world, and the privacy-enhanced environment.
In addition, this section provides a deep analysis of how our proposed attack works in IOTA 2.0.

\subsection{Decentralized environment}
In a decentralized environment, we assume a situation in which IOTA is completely decentralized.
As control variables, $N$, the total number of full nodes, $p$, the ratio of $C$ and $N$, and $M$, the number of nodes requesting tip selection, were set. For their default values, $N$ is set to 100, $p$ is set to 0.1, and $N_s$ is set to 3, which is the value used in the IOTA client, PyOTA \cite{url3}. 
Changes in $N$ and $M$ have no effect, and as $p$ increases, the deanonymization probability $P(A|B_i)$ increases.
In particular, as $p$ and $P(A|B_i)$ are calculated to be similar values, we can say $p\approx P(A|B_i)$.

\subsection{Real world node distribution}

In the real world, it is hard to guarantee that every full node has an equal chance of being chosen when a light node requests a tip selection.
Requesting information from full nodes that are far away from the main relationship is a waste of time and they might not have the address information that the light node needs.
Therefore, we analyze the distribution of actual IOTA full nodes and consider all possible outcomes.

\begin{figure}[tb]
    \centering
    \includegraphics[width=\linewidth]{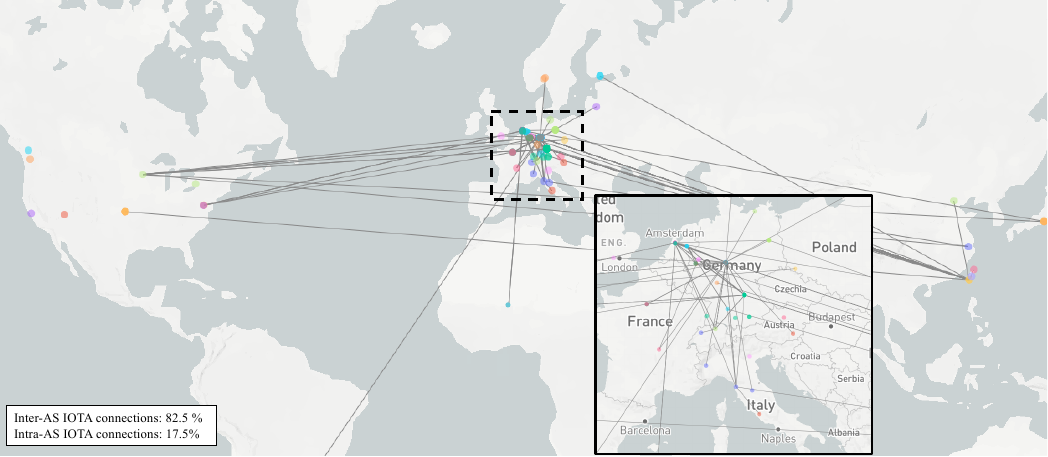}
    \caption{Global map of IOTA full nodes with neighboring relationships.\cite{ref3}}
    \label{fullnode}
\end{figure}

Fig.~\ref{fullnode} displays the distribution of IOTA full nodes.
It shows that only a few full nodes exist on non-European continents.
In Asia, for instance, if one malicious full node is added based on the above, anonymity attacks can be performed on 14\% of transactions.
If all nodes in China collude, 83\% of transactions in Asia can be deanonymized.

Let us explore this topic more thoroughly.
When a light node requests tip selection to register a transaction, it does so from a full node within the same continent for efficiency reasons.
As per Fig. \ref{fullnode}, there are 8 full nodes in North America, 31 in Europe, 1 each in South America and Africa, and 6 in Asia. 
The adversary can take control of each node, which can lead to deanonymization.
For instance, nodes in North America can be deanonymized for 1/8=12.5\% of transactions for every node taken over.
Similarly, for nodes in Europe, deanonymization is possible for 1/31$\approx$3.2\% of transactions when one node is taken over. 
There is a 16.7\% chance of an attack on transactions for every full node in Asia.
In South America and Africa, if the node is an adversary node, the neighboring light nodes are connected to the full node by default, which can cause deanonymization issues for all transactions.
To achieve the same level of deanonymization as in North America, we need to have four adversary nodes for every one full node in Europe.
This is due to the small number of full nodes and their uneven distribution, which makes the attack easier to carry out.

\begin{figure}[tb]
    \centering
    \includegraphics[width=0.9\linewidth]{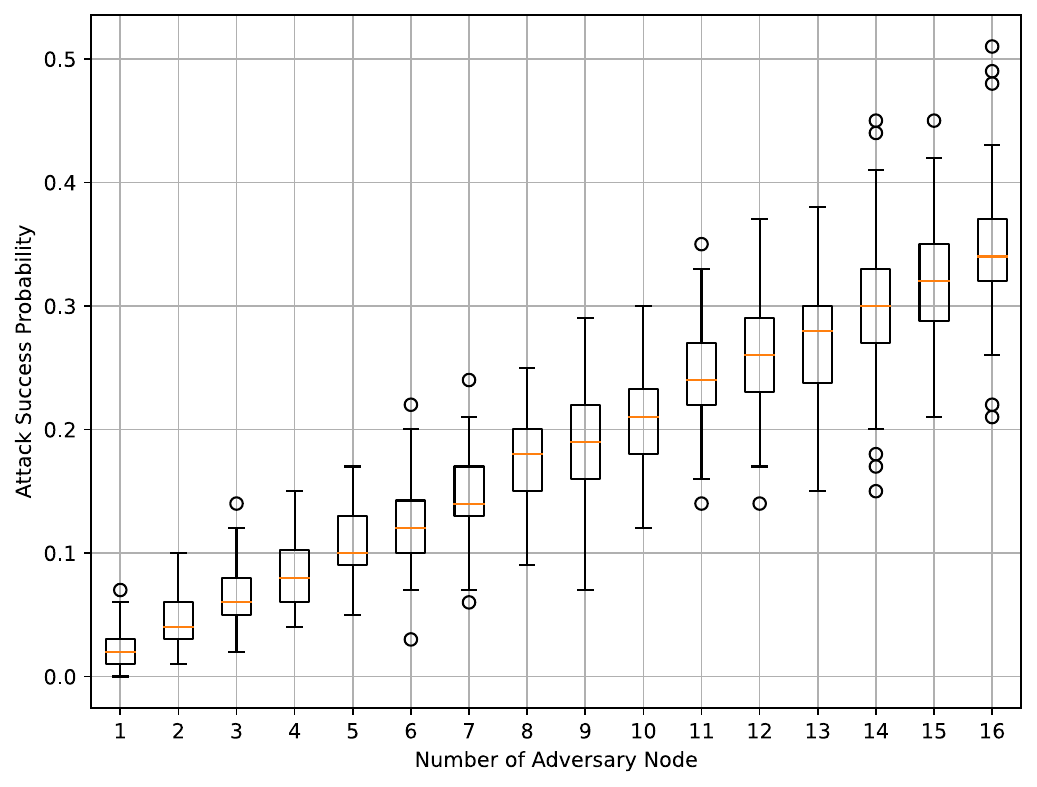}
    \caption{Attack Success Probability in the Real World}
    \label{box}
    \vspace{-0.3cm}
\end{figure}

The graph in Fig.~\ref{box} displays the probability of success of our proposed attack when adversary nodes are selected from the real-world full node distribution.
The number of adversary nodes is limited to 16 since distributed computing assumes that more than 1/3 of faults do not guarantee correct system behavior.
To account for numerous cases, we randomly selected as many adversary nodes as there were, up to 100 samples.
For each sample, we calculated the probability of the light node deanonymization.
Compared to the decentralized environment, the average probability of attack success is similar, but depending on the distribution of adversaries, higher attack success rates are possible.

To gain a clear understanding of the effects of centralization, we designed an experiment.
We abstract the real world into a single plane and randomly distributed full nodes.
We divide the distribution into nine grids, and for each grid, we create a heatmap illustrating the probability of an adversary being selected when a light node requests tip selection.

In this experiment, the total number of nodes was set to 50, the size of the plane was 10 by 10, the adversary ratio was 0.1, and light nodes were allowed to request tip selection only from full nodes within a 3-unit radius.
To emphasize the effects of centralization, we ensured that at least one adversary node was present in the same grid.

\begin{figure*}
    \centering
    \begin{subfigure}[tb]{0.49\textwidth}
        \centering
        \includegraphics[width=0.9\linewidth]{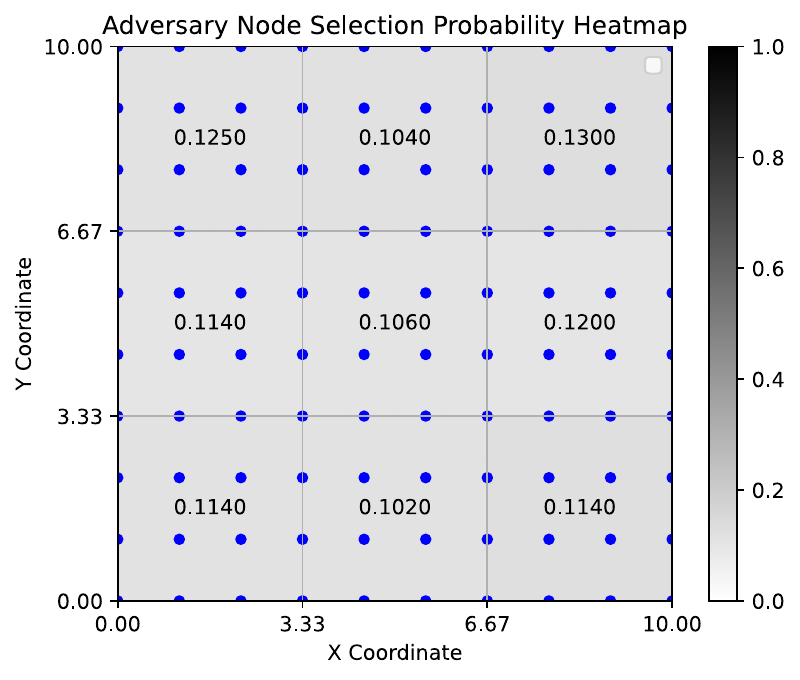}
        \caption{Adversary selection probability heatmap in uniform distribution}
        \label{uniform}

    \end{subfigure}
    \hfill
    \begin{subfigure}[tb]{0.49\textwidth}
        \centering
        \includegraphics[width=0.9\linewidth]{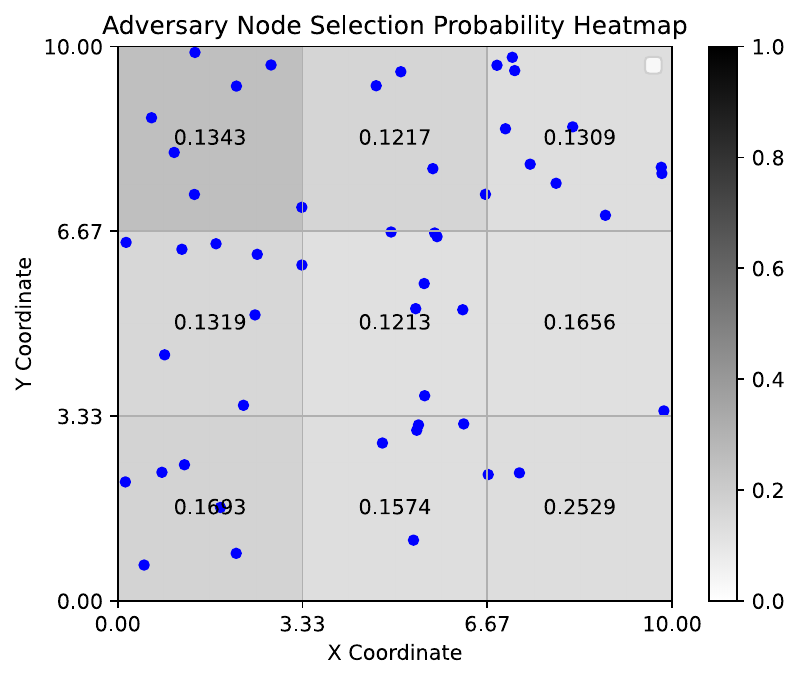}
        \caption{Adversary selection probability heatmap in random distribution}
        \label{random}
    \end{subfigure}
    \caption{Adversary selection probability heatmap}
\end{figure*}

The probability of an adversary receiving tip selection requests remains relatively stable across most grids in a uniformly distributed scenario, as shown in Fig.~\ref{uniform}. 
However, in a randomly distributed scenario, as illustrated in Fig.~\ref{random}, full nodes tend to cluster in specific grids. 
This can be problematic as having only a few adversary nodes in these regions can be enough for deanonymization.
This issue is particularly noticeable in scenarios where there is a scarcity of nodes. 
In grids with a significant concentration of full nodes, the probability of an adversary receiving tip selection requests is not significantly different from the theoretical value of $p=0.1$. 
However, in the area with fewer full nodes, a probability of 0.2529 was observed.

\begin{figure}[tb]
    \centering
    \includegraphics[width=0.9\linewidth]{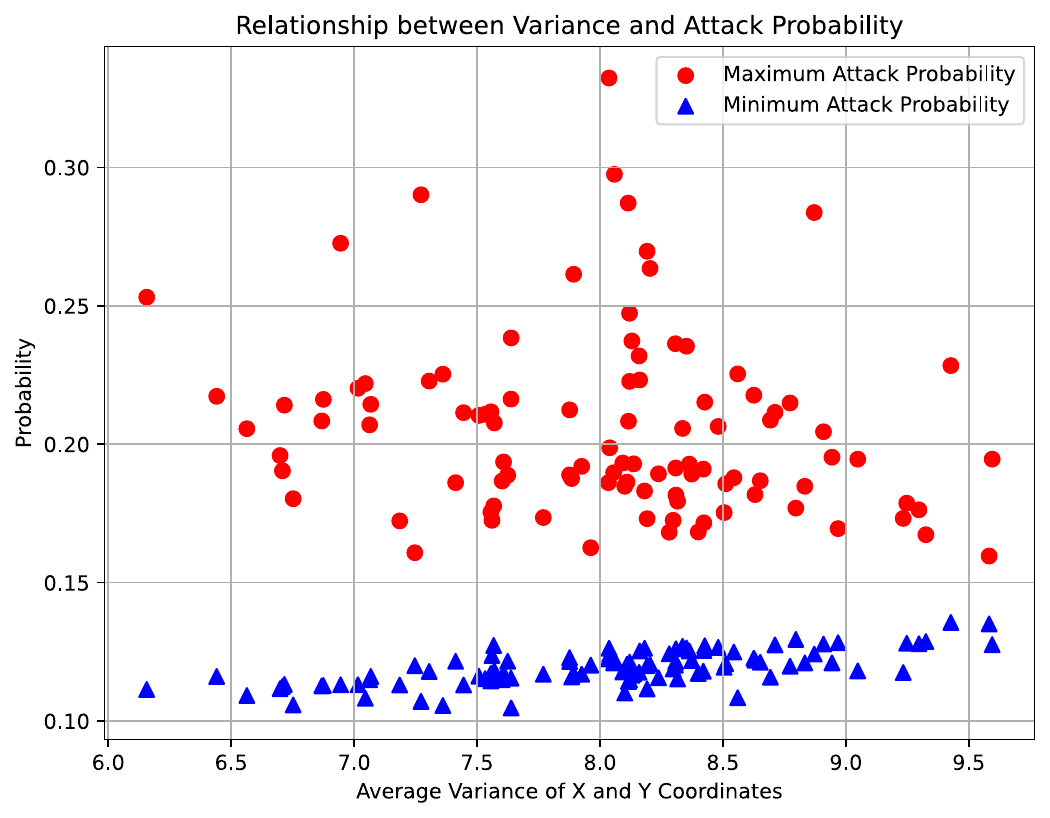}
    \caption{Maximum and minimum adversary selection probability in random distribution (N=100)}
    \label{variance}
\end{figure}

After conducting multiple simulations with random distributions, we have discovered that any distribution can lead to the same outcome.
In Fig.~\ref{variance}, we present a graph that illustrates the relationship between the variance and the minimum and maximum probability values in the grid.
We generated a random distribution and ran the simulation 100 times. 
The minimum probability value indicates that as the variance increases, the number of nodes placed on the same grid decreases, increasing the probability of an adversary being selected.
On the other hand, the maximum probability value does not show a trend when the number of simulations is 100.
However, we can observe that the probability of an adversary being selected increases from a minimum of 4\%p to a maximum of over 20\%p in grids with fewer nodes than the minimum probability.

In reality, IOTA's centralization contributes to this phenomenon.
We propose IOTA's decentralization and deployment of more full nodes as a course of action.
The deployment of more full nodes will decrease the $p$ value, resulting in a decrease in the probability of a deanonymization attack itself.
In addition, less centralization not only resolves various problems currently in IOTA but also leads to a decrease in the $p$ value due to an increase in the number of full nodes that can request tip selection from light nodes.
The number of recommended full nodes varies depending on the intensity of anonymity to be held.

\subsection{Privacy-enhanced environment}
The proposed methodologies described in Section II primarily focus on aspects related to taint analysis and the topology of full or light nodes.
We explore the feasibility of applying tip selection attacks to methodologies that can be employed within the IOTA and are resistant to taint analysis.
However, it is important to note that taint analysis involves tracking transaction histories.
In contrast, the tip selection attack operates at the transaction initiation stage without prior knowledge, and these mentioned methodologies may not be sufficient to prevent the attacks we have introduced.

In the decentralized mixing proposed by Sarfraz et al.\cite{ref17}, $n$ peers each possess $n$ inputs and participate in the mixing process, which results in the registration of transactions on the ledger.
In this scheme, peers generate addresses with no transaction history or balance and utilize these addresses as intermediaries to carry out transactions.
Consequently, even if an adversary were to obtain a mapping of addresses to individuals through a tip selection attack, the adversary would not have access to the actual contents of the executed transactions.

However, the methodology presented in this paper focuses on mapping addresses to individuals, rather than the content of transactions, with an emphasis on the information of peers participating in mixing.
Consider $n$ peers engaging in transactions in IOTA using a mixer.
As $n$ transactions occur, $n$ tip selection requests are made, and these requests can be probabilistically associated with $N$ by a ratio of $C$, denoted as $p$.
Subsequently, when utilizing newly generated addresses for transactions, there is also a probability of $p$ that mapping between addresses and individuals becomes possible, resulting in a probability of $p^2$ for mapping between input and output addresses.

In an environment where a mixer is utilized for all transactions, mapping between addresses and individuals may not reveal significant information.
When $N$ peers initially use the mixer, we can determine $E(x)=np$ address-individual mappings, but cannot identify that all $np$ mappings belong to the same mixer transaction participants.
Due to the characteristics of the mixer, identifying peers by finding the transaction associated with the initially used address is also meaningless.

Suppose participants $\{a_1 ... a_n\}$ of a mixer transaction, meaning a transaction $\{a_1...a_n\}\rightarrow\{b_1...b_n\}$ occurs.
Let us define a function $y=f(x)$ to represent an individual $y$ associated with address $x$.
We can say that $a_1$ and $b_1$ are addresses involved in the same transaction.
However, since $b_1$ is an initially used address, we have no information to determine who owns this address.
Therefore, we have to ascertain the owner of $y_2=f(b_1)$ with the probability of $p$. Consequently, $a_k$, the original address of $y_2$, also requires a probability of $p$ for a successful mapping.
Thus, we can identify $y_2$, which participated in the same transaction as $y_1$ with the probability of $p^2$.

To put it more explicitly, to know that $a_i$ and $a_j$ are the same mixer transaction participants, it is necessary to determine $f(a_j) = f(b_i)$.
Without loss of generality, when $y_1=f(a_1)$ is known, the following steps are required to identify the same transaction participants:
\begin{equation}
    \begin{split}
        a_1\rightarrow b_1\ transaction: P(f(a_i)=f(b_1)) = p^2 \\
        a_i\rightarrow b_i\ transaction: P(f(a_j)=f(b_i)) = p^2 \\
        a_j\rightarrow b_j\ transaction: P(f(a_k)=f(b_j)) = p^2 \\
        \vdots\qquad\qquad\qquad\qquad\qquad
    \end{split}
\end{equation}
Such steps must occur in succession, and we can calculate the probability of identifying $x$ transaction participants as $P(x) = p^{2(x-1)}$.
Calculating the expected value of x results in $E(x) = \sum_i^{\infty}iP(i) = \sum_i^{\infty}ip^{2(i-1)}$. 
When $p=0.1$, the approximate value is $E(x)\approx1.01$. This indicates that, in practice, only 1.01 participants can be identified, demonstrating that it is difficult to identify mixer transaction participants.

\subsection{IOTA 2.0}
IOTA is currently developing IOTA 2.0 due to a centralization problem caused by coordinators\cite{url1}.
It is expected to resolve centralization through the introduction of \textit{coordicide} and is currently in the stage of verifying its effectiveness with the IOTA 2.0 Devnet\cite{ref10}.
IOTA 2.0 is different from IOTA in that it does not rely on a coordinator to function.
Instead of achieving the ledger's agreement through communication between nodes, nodes observe the Tangle and reach an agreement.

Despite adopting IOTA 2.0, the issues outlined in this paper are unlikely to be resolved.
Even after the removal of the coordinator, IOTA 2.0 will still need to choose references from a pool of eligible tips due to Tangle's inherent nature.
This tip selection algorithm is difficult to execute on light nodes and will ultimately require the use of full nodes, rendering our attack still viable.
Furthermore, the elimination of the coordinator is not anticipated to significantly alter the current distribution of full nodes.
Nonetheless, the introduction of sharding, which involves breaking down large databases into smaller segments, is expected to enhance transaction throughput and address the scalability problem, increasing the number of full nodes. This could potentially alleviate the issue to some extent.

\section{Mitigation Against the Proposed Attack}
\label{section: mitigation}
As shown in Section \ref{section: analysis}, the attack is valid in the existing environment due to the dependency of light nodes on full nodes to perform tip selection. In this section, we introduce a mitigation against our proposed attack.

\subsection{Full node scaling and global tip selection request}
When the tip selection is requested globally and the number of full nodes increases, our proposed attack's success rate will decrease, even in regions where the number of neighboring full nodes is low.
The success rate of our attack depends on the ratio of adversary nodes to entire full nodes.
In regions with fewer neighboring full nodes, we had a higher attack success rate with only a small number of adversary nodes because fewer full nodes could receive tip selection requests.
However, as the number of full nodes increases and tip selection requests become global, more full nodes can be requested in all regions.
As a result, the attacker will require more adversary nodes to achieve the original attack success rate.

As previously mentioned, the success of our attack hinges on the ratio of all nodes to the adversary node. 
To effectively execute this method, we must first determine the strength of anonymity desired, as well as the capabilities of the attacker.
We must know how many adversary nodes the attacker can operate before proceeding. 
Once this is established, we can then calculate the number of full nodes required to match the desired level of anonymity. 
For instance, if the attacker can operate up to 10 adversary nodes and we expect the attack success rate to remain below 1\%, we should have over 1000 operational full nodes, with light nodes capable of requesting tip selection to all full nodes.

However, there are several issues with this approach.
First of all, the increase in the number of full nodes is difficult to resolve by IOTA itself.
Increasing the number of full nodes operated by IOTA goes against the nature of a distributed system.
As IOTA's market expands, the number of full nodes should naturally increase.
Secondly, when a tip selection is requested from all full nodes, communication between full nodes is unavoidable. 
Without appropriate encapsulation, unwanted full nodes may know the contents of the tip selection request.
To address this, encrypting the tip selection request with the public key of the receiving full node may be a suitable solution.

\subsection{Introducing proxy nodes}
A proxy node between the light and full nodes can be another solution. Let a proxy node be an intermediary between a light node and a full node.
It fetches data from the full node on behalf of the light node, preventing direct connection between them, and providing only the necessary information.
The main goal of the proxy node is to maintain anonymity while processing requests from the light node. In this way, the light node can avoid direct communication with the full node and need not be concerned about the minute details being revealed during the transfer of information.

We propose two conditions to implement a proxy node.
Firstly, efficient data management is required. Proxy nodes should utilize caching and data compression to ensure fast data delivery and maintain network efficiency.
This minimizes the bandwidth required for data transmission while maintaining security and low latency. 
Secondly, the implementation of a proxy node should be designed for network scalability.
This is necessary to cope with the increasing transaction load and maintain the efficiency of the entire network.

The communication between the light node and the full node is indirect in this method.
This means that the adversary cannot gather metadata such as the IP of the light node.
By comparing the ledger with the requested tip selection, the adversary may know which proxy node the transaction came from, but it is very challenging to identify the light node that is connected to that proxy node.
However, it is essential to note that this solution assumes a trusted proxy node.
If an adversary operates a proxy node, it may cause more privacy concerns than before.
Furthermore, communicating through the proxy node may result in a delay issue, which may reduce the effectiveness of the solution.

\subsection{Direct tip selection by light nodes}
In a previous explanation, we mentioned that light nodes rely on full nodes for tip selection due to performance issues.
However, there is still work to be done to address these performance issues.
Light nodes cannot manage the entire ledger as it is difficult to run them around the clock and they lack the necessary storage capacity.

To address this issue, we propose that light nodes manage only the minimum amount of recent transaction information required for tip selection.
The light node would periodically receive recent transaction information from their connected full node to perform this.
Additionally, a simplified tip selection algorithm should be implemented that selects eligible tips based on recent transaction information.
Once these two issues are addressed, light nodes will be able to perform tip selection directly.

The light node in this case directly performs the tip selection without communicating with any other full node except the one it is connected to.
As a result, it is impossible to compare the communication with the ledger, and the attack fails.
The adversary node can only determine which full node registers the transaction in the ledger.
However, the effectiveness of this solution depends heavily on the performance of the light node.
If the light node cannot maintain a recent ledger or if it does not have enough memory to perform the tip selection algorithm, the scheme will fail.

\subsection{Effectiveness of mitigations}

\begin{table}[tb]
    \caption{Comparison of each environment against proposed attack}
    \centering
    \renewcommand{\arraystretch}{2}
    \begin{tabular}{p{2.7cm} ccc}
        \textbf{Environment} & \textbf{Resistance} & \textbf{Latency} & \textbf{Decentralization} \\ \hline
        Real world & \xmark & \cmark & $\Delta$ \\
        Privacy-enhanced & $\Delta$ & $\Delta$ & $\Delta$ \\
        IOTA 2.0 & \xmark & \cmark & \cmark \\
        Full node scaling & $\Delta$ & \cmark & $\Delta$ \\
        Proxy node & \cmark & $\Delta$ & \xmark \\
        Light node tip selection & \cmark & \xmark & $\Delta$ \\
    \end{tabular}
    \footnotesize{\cmark: fully provided, $\Delta$: partially provided, \xmark: not provided}
    \label{comparison_env}
\end{table}

We compare the effectiveness of our attack in different environments and evaluate the appropriate response to it based on three important properties: resistance, latency, and decentralization.

Table \ref{comparison_env} presents a comparison of each environment against the proposed attack.
Our proposed attack can deanonymize the light node, which can cause serious damage.
Therefore, resistance is the most important property we need to consider.

Our proposed attack is not resisted in the real world and IOTA 2.0 environments.
However, in a privacy-enhanced environment, it is harder to determine the meaning of the obtained information, which requires additional effort.
Meanwhile, in the environment with the two mitigations we proposed, our attack was not successful.

While resisting attacks, it is also important to maintain low latency and decentralization in line with IOTA's ethos.
The real world environment faces a challenge in decentralization due to the presence of coordinators.
The introduction of IOTA 2.0 is expected to achieve decentralization to some extent.
In the privacy-enhanced environment, there may be some latency when performing decentralized mixing, and centralization is not solved as the introduction of this method does not affect the full node distribution.

Full node scaling and global tip selection requests can reduce the effectiveness of our attack.
Although requesting tip selection globally may cause a delay while creating a transaction, it will not impact the time taken to register the transaction, which means it will not have a significant effect on the overall IOTA experience.
However, increasing the number of full nodes reduces the latency.
While it does not directly impact decentralization, global tip selection requests can lead to a decentralized effect compared to the current environment.

Light node tip selection and the introduction of proxy nodes can resist our attack, but additional work is needed in terms of latency and decentralization.
In the case of light nodes performing tip selection, decentralization can be achieved with the introduction of IOTA 2.0, but problems may arise due to latency in the process of performing tip selection.
The introduction of proxy nodes has less latency than light node tip selection, but the introduction of trusted proxy nodes can accelerate centralization.

\section{Discussion}
\label{section: discussion}
This section introduces the significance of our attack and outlines the limitations of our analysis, as well as our plans.

\subsection{Significance of our attack}
Bonneau et al.\cite{ref22} outline four primary techniques for deanonymizing cryptocurrency.
Two of these techniques, taint analysis, and transaction analysis, were previously introduced in section \ref{subsection: privacy}.
The third technique involves side-channel attacks, such as timing and precise value.
The final technique is to link pseudonyms and identities by using auxiliary information that the attacker gleans from forums or merchants.

The first three techniques rely on the fourth one to link a user to a real-world identity.
However, we suggest another way that compares tip selection request-response and ledger data instead of combining external information.
Our proposed attack aims to match the address with a user's metadata, which would make it easier to determine the identity of multiple addresses based on the address associations acquired through the other techniques.

\subsection{Full node distribution}
\label{subsection: full node distribution}
The full node distribution data used in this paper is from 2020, and there is currently no accurate data on the current distribution of full nodes.
Therefore, it is necessary to measure the risk in the real world through additional IOTA topology analysis.
Perazzo et al.\cite{ref3} analyzed the entire topology using a recursive algorithm that queried the entire node for a neighbor list.
Since IOTA does not provide information on the full node distribution, the additional analysis should be performed after network configuration using the same method as above.

\subsection{Tip selection algorithm}
\label{subsection: tip selection algorithm}
In this paper, when analyzing the problem of deanonymization in the tip selection process, we assumed that a different response was sent by each full node to all tip selection requests.
However, in reality, different requests can produce identical responses, and it is necessary to distinguish whether the message was affected by the response of the adversary or if it accidentally generated the same tip selection for another request.
It is only possible to accurately calculate the probability of deanonymization by taking additional considerations into account.

IOTA does not provide a standard for tip selection algorithm but recommends Uniform Random Tip Selection (URTS) or Markov Chain Monte Carlo (MCMC) methods\cite{ref1}.
In particular, since the introduction of IOTA 2.0, URTS has been used as the main algorithm.
Taking this into account, an accurate deanonymization probability can be calculated through additional calculations on the probability that the same result can be obtained due to URTS.
Since the URTS algorithm utilizes random choices, conditional probabilities are expected to be derived using Bayes' theorem.

\section{Conclusion}
\label{section: conclusion}

In this paper, we have presented a deanonymization attack on IOTA, which takes place during the tip selection phase.
This attack takes place when a light node requests a full node due to performance limitations, and the full node can determine the origin of the transaction by comparing the response with the ledger.
Our research highlights the practicality of such attacks within the current IOTA framework and emphasizes their persistence in IOTA 2.0 and ongoing privacy enhancements.
We have also proposed effective strategies to mitigate these vulnerabilities, to strengthen privacy without compromising the network's efficiency and decentralization.
In future work, we plan to investigate whether similar vulnerabilities exist in other cryptocurrencies that use DAG, such as Obyte and NANO. 
As well as advanced attack concepts, mitigation methods to maintain efficiency and scalability should be studied against tip selection deanonymization.

\section*{Acknowledgement}
 
This work was supported by Institute of Information \& communications Technology Planning \& Evaluation (IITP) grant funded by the Korea government(MSIT) (No.2021-0-00613, Zero Trust technology based access control and abnormal event analysis technology development for enterprise network protection in the untact era)

\bibliographystyle{ieeetr}
\bibliography{paper.bbl}

\end{document}